\documentclass[preprint,12pt]{elsarticle}
\usepackage{amssymb}

\journal{Applied Mathematics and Computation}

\begin{document}

\begin{frontmatter}

\title{Approximations to the solution of Cauchy problem for a linear evolution equation via the space shift operator\\ (second-order equation example)}

\author{Ivan\,D.\,Remizov}

\address{Bauman Moscow State Technical University\\
Rubtsovskaya nab., 2/18, office 1027, 105005 Moscow, Russia\\
Lobachevsky Nizhny Novgorod State University\\
Prospekt Gagarina, 23, Nizhny Novgorod, 603950, Russia}

\ead{ivan.remizov@gmail.com}

\begin{abstract}
We present a general method of solving the Cauchy problem for a linear parabolic partial differential equation of evolution type with variable coefficients and demonstrate it on the equation with derivatives of orders two, one and zero. The method is based on the Chernoff approximation procedure applied to a specially constructed shift operator. It is proven that approximations converge uniformly to the exact solution.
\end{abstract}


\end{frontmatter}

\begin{center}
\textbf{Problem setting and approach proposed}
\end{center}

Consider $x\in\mathbb{R}^1$, $t\geq0$ and set the Cauchy problem for a second-order parabolic partial differential equation
$$
\left\{ \begin{array}{ll}
   u'_{t}(t,x)=(a(x))^2u_{xx}''(t,x)+b(x)u'_x(t,x) +c(x)u(t,x)=Hu(t,x),  \\
   u(0,x)=u_0(x).
\end{array} \right.\eqno(1)
$$
The coefficients $a$, $b$, $c$, $u_0$ above are bounded, uniformly continuous functions $\mathbb{R}^1\to\mathbb{R}^1$. This paper is dedicated to deriving of an explicit formula that expresses the solution of (1) in terms of $a$, $b$, $c$, $u_0$ assuming that the operator $H$ is an infinitesimal generator of the $C_0$-semigroup $\left(e^{tH}\right)_{t\geq 0}$. This assumption is a standard one in studies of evolution equations, the class of equations that considered equation belongs to. According to the general theory of $C_0$-semigroups \cite{EN1} this assumption implies that the solution of the Cauchy problem (1) exists, is bounded and uniformly continuous with respect to $x$ for each $t$, depends on $u_0$ continuously and can be represented in a form $u(t,x)=\left(e^{tH}u_0\right)(x)$. We apply the Chernoff theorem \cite{Chernoff} to a specially constructed family of operators $(S(t))_{t\geq 0}$, and express $e^{tH}$ in terms of $a$, $b$, $c$ reaching the goal announced. We do not discuss the problem of finding the class of functions in which the solution is unique under certain assumptions on functions $a$, $b$, $c$, $u_0$, but keep in mind that e.g. for a heat equation there are known unbounded solutions. 

The formula that provides the solution of (1) is given in theorem 3.

\begin{center}
\textbf{Technique employed} 
\end{center}
The Chernoff theorem \cite{EN1, Chernoff} allows to reduce the problem of finding $e^{tH}$ to the problem of finding an appropriate operator-valued function $S(t)$, which is called the Chernoff function, and then use the Chernoff formula $e^{tH}=\lim_{n\to\infty}S(t/n)^n$. One advantage of that step is that we can define $S(t)$ by an explicit formula that depends on the coefficients of the operator $H$. Members of O.G.Smolyanov's group employed this technique using integral operators as Chernoff functions to find solutions to parabolic equations in many cases during last 15 years, see the pioneering papers  \cite{SWW, STT}, overview \cite{SmHist}, introduction to \cite{R1} and two more exotic examples \cite{Dubravina, SmSh}. The solutions obtained there were represented in the form of \textit{Feynman formula}, i.e. as a limit of a multiple integral as the multiplicity goes to infinity. The Schr\"odinger equation also belongs to the class of evolution equations, and the same way allows to represent the Cauchy problem solution for it in the form of Feynman and quasi-Feynman integral formulas, see \cite{R2} and references therein.

The specific feature of the research that is now presented is that we use the shift operators instead of integral operators when constructing the Chernoff function $S(t)$. For this reason the solution of (1) is now represented via a new type of formulas that do not include integrals. Our model example, i.e. the Cauchy problem (1), can be modified in different directions. The example was chosen rather simple (free of distinctive or complicated details) intentionally to let the reader focus on the the main contents of the article i.e. on the presented method. 

\textbf{Definition 1.} Let $\mathcal{L}(\mathcal{F})$ be the set of all linear bounded operators in a Banach space $\mathcal{F}$. Let the operator $L\colon \mathcal{F}\supset Dom(L)\to \mathcal{F}$ be linear and closed. The function $G$ is called \textit{Chernoff-tangent} to the operator $L$ iff:

(CT1) $G$ is defined on $[0,+\infty)$, takes values in $\mathcal{L}(\mathcal{F})$, and the function $t\longmapsto G(t)f$ is continuous for each $f\in\mathcal{F}$. 

(CT2) $G(0)=I$, i.e. $G(0)f=f$ for each $f\in\mathcal{F}$.

(CT3) There exists such a dense subspace $\mathcal{D}\subset \mathcal{F}$ that for each $f\in \mathcal{D}$ there exists a limit $$G'(0)f=\lim_{t\to 0}\frac{G(t)f-f}{t}.$$ 

(CT4) The closure of the operator  $(G'(0),\mathcal{D})$ is equal to $(L,Dom(L)).$ 

\textbf{Remark 1.} In the definition of the Chernoff tangency the family $(G(t))_{t\geq 0}$ usually does not have a semigroup composition property. However, each $C_0$-semigroup is Chernoff-tangent to its generator.

\textbf{Theorem 1} (\textsc{P.\,R.~Chernoff, 1968}). Let $\mathcal{F}$ and $\mathcal{L}(\mathcal{F})$ be as before. Suppose that the operator $L\colon \mathcal{F}\supset Dom(L)\to \mathcal{F}$ is linear and closed, and function $G$ takes values in $\mathcal{L}(\mathcal{F})$. Suppose that these assumptions are fulfilled:

(E) There exists a $C_0$-semigroup $(e^{tL})_{t\geq 0}$ with the generator $(L,Dom(L))$.

(CT) $G$ is Chernoff-tangent to $(L,Dom(L)).$ 

(N) There exists such a number $\omega\in\mathbb{R}$, that $\|G(t)\|\leq e^{\omega t}$ for all $t\geq 0$.

Then for each $f\in \mathcal{F}$  we have $(G(t/n))^nf\to e^{tL}f$ as $n\to \infty$ with respect to norm in $\mathcal{F}$ uniformly with respect to $t\in[0,t_0]$ for each $t_0>0$.

\textbf{
\begin{center}
Main result
\end{center}
}
\textbf{Remark 2.} The main result of the paper is formula (4) proven in theorem~3. It contains $\lim_{n\to\infty}$. After the limit is taken we obtain the exact solution to Cauchy problem (1). For each fixed $n$ the expression under the limit sign is an approximation to the solution. With growth of $n$ such approximations converge to the exact solution uniformly with respect to $x\in\mathbb{R}^1$ and $t\in[0,t_0]$ for each fixed $t_0>0$.

\textbf{Remark 3.} Let us denote the set of all (real-valued and defined on the real line) bounded continuous functions as $C_b(\mathbb{R})$, the set of all bounded functions with bounded derivatives of all orders as $C^{\infty}_b(\mathbb{R})$, and the set of all bounded, uniformly continuous functions as $UC_b(\mathbb{R})$. 

Then $C^{\infty}_b(\mathbb{R})\subset UC_b(\mathbb{R})\subset C_b(\mathbb{R})$, and with respect to the uniform (Chebyshev) norm $\|f\|=\sup_{x\in\mathbb{R}}|f(x)|$ the first inclusion is dense, and the last two spaces are Banach spaces. 

\textbf{Theorem 2.} For each $x\in\mathbb{R}$, $t\geq 0$, $f\in C_b(\mathbb{R})$ and $\varphi\in C^{\infty}_b(\mathbb{R})$ set
\begin{flushleft}
\small $(S(t)f)(x)=\frac{1}{4}f\left(x+2a(x)\sqrt{t}\right)+\frac{1}{4}f\left(x-2a(x)\sqrt{t}\right)+\frac{1}{2}f(x+2b(x)t)+tc(x)f(x),\ $\normalsize (2)
\end{flushleft}
$$(H\varphi)(x)=(a(x))^2\varphi''(x)+b(x)\varphi'(x)+c(x)\varphi(x).\eqno(3)$$
Then with respect to the norm $\|g\|=\sup_{x\in\mathbb{R}}|g(x)|$ the following hold:

I) for each $t\geq 0$ and $f\in C_b(\mathbb{R})$ we have $\|S(t)f\|\leq \big(1+\|c\|t\big) \|f\|$.

II) for each $\varphi\in C^{\infty}_b(\mathbb{R})$ we have $\lim_{t\to+0}\|S(t)f-f-tHf\|/t=0$.

III) if $t_n\to t_0$, $t_n\geq 0$ and $f\in UC_b(\mathbb{R})$ then $\lim\limits_{t\to t_0}\|S(t_n)f- S(t_0)f\|=0$ for each $t_0\geq 0$.
 
IV) if $a,b,c,f\in UC_b(\mathbb{R})$ then $S(t)f\in UC_b(\mathbb{R})$ for each $t\geq 0$.

\textbf{Proof.} Let us write $\sup$ instead of $\sup_{x\in\mathbb{R}}$ to make it shorter.  

I) $\|S(t)f\|=\sup|\frac{1}{4}f(x+2a(x)\sqrt{t})+\frac{1}{4}f(x+2a(x)\sqrt{t})+\frac{1}{2}f(x+2b(x)t)+tc(x)f(x)|\leq  \frac{1}{4}\sup|f(x+2a(x)\sqrt{t})|+\frac{1}{4}\sup|f(x+2a(x)\sqrt{t})|+\frac{1}{2}\sup|f(x+2b(x)t)|+t\sup|c(x)|\sup|f(x)|\leq \frac{1}{4}\|f\|+\frac{1}{4}\|f\|+\frac{1}{2}\|f\|+t\sup|c(x)|\|f\|=\big(1+\|c\|t\big) \|f\|.$

II) Let us fix arbitrary $x\in\mathbb{R}$ in (2). Let us use Taylor's formula and expand the first two summands in (2) in powers of $\sqrt{t}$, and the third summand in powers of $t$; let us represent remainders in Lagrange's form:
\small
$\varphi(x+2a(x)\sqrt{t})=\varphi(x)+ 2a(x)\sqrt{t}\varphi'(x) + \frac{1}{2}\varphi''(x)(2a(x)\sqrt{t})^2 + \frac{1}{6}\varphi'''(\xi_1)(2a(x)\sqrt{t})^3$,\\
$\varphi(x-2a(x)\sqrt{t})=\varphi(x)- 2a(x)\sqrt{t}\varphi'(x) + \frac{1}{2}\varphi''(x)(2a(x)\sqrt{t})^2 - \frac{1}{6}\varphi'''(\xi_2)(2a(x)\sqrt{t})^3$,\\
$\varphi(x+2b(x)t)= \varphi(x)+2b(x)t\varphi'(x)+\frac{1}{2}\varphi''(\xi_3)(2b(x)t)^2$. 
\normalsize

Using this and (2) we write an expression for $(S(t)f)(x)$ and then transform it using (3). Next we majorize $\sup|(S(t)f)(x)-f(x)-tHf(x)|$ keeping in mind that functions $a,b$ are bounded, and derivatives of the function  $\varphi\in C^{\infty}_b(\mathbb{R})$ are bounded. It leads us to

$(S(t)\varphi)(x)=\varphi(x)+t\big[(a(x))^2\varphi''(x)+b(x)\varphi'(x)+c(x)\varphi(x)\big]+o(t)$.

III) The function $a$ is bounded, so $x+2a(x)\sqrt{t_n}\to x+2a(x)\sqrt{t_0}$ uniformly with respect to $x$. Function $f$ is uniformly continuous, so $f(x+2a(x)\sqrt{t_n})\to f(x+2a(x)\sqrt{t_0})$ uniformly with respect to $x$. The rest is obvious.

IV) If $t\geq 0$ is fixed, then $\big[z\longmapsto f(z+2a(z)\sqrt{t})\big]\in UC_b(\mathbb{R})$ because  $a,f\in UC_b(\mathbb{R})$. The rest is obvious.

\textbf{Remark 4.} The above proof is true for functions $N\to\mathbb{R}$ on arbitrary (finite- or infinite-dimensional) normed space $N$. We only need to substitute $C^{\infty}_b(\mathbb{R})$, $UC_b(\mathbb{R})$, $C_b(\mathbb{R})$ by $C^{\infty}_b(N,\mathbb{R})$, $UC_b(N,\mathbb{R})$, $C_b(N,\mathbb{R})$ and consider derivatives in the sense of Fr\'echet. 

\textbf{Theorem 3.} Suppose that functions $a$, $b$, $c$ belong to the space $UC_b(\mathbb{R})$ endowed with the norm $\|f\|=\sup_{x\in\mathbb{R}}|f(x)|$. Suppose that operator $H$ is defined by equality (3) on the domain $C^\infty_b(\mathbb{R})\subset UC_b(\mathbb{R})$, and the closure of this operator: a)~exists;
b)~is~an~infinitesimal generator of a $C_0$-semigroup $(e^{tH})_{t\geq0}$ in $UC_b(\mathbb{R})$.

Then for each $u_0\in UC_b(\mathbb{R})$ there exists a bounded (and uniformly continuous with respect to $x\in\mathbb{R}$ for each $t\geq 0$) solution $u$ of the Cauchy problem~(1), it depends on $u_0$ continuously and uniformly with respect to $x\in\mathbb{R}$ for each $t\geq0$. For each $x\in\mathbb{R}$ and $t\geq0$ this solution is given by the formula $$u(t,x)=\left(e^{tH}u_0\right)(x)=\lim_{n\to\infty}\Big(\Big(S(t/n)\Big)^nu_0\Big)(x),\eqno(4)$$
where $S(t/n)$ is obtained by substitution of $t$ by $t/n$ in the equality (2), and the $n$-th power means the composition of $n$ copies of linear bounded operator $S(t/n)$. The limit (4) for each fixed $t>0$ is taken in the space $UC_b(\mathbb{R})$ and appears to be uniform with respect to $t\in[0,t_0]$ for each $t_0>0$.

\textbf{Proof.} Let us check the conditions of the Chernoff theorem. In theorem~1 and definition 1 we set $\mathcal{F}=UC_b(\mathbb{R})$, $G(t)=S(t)$, $L=H$, $\mathcal{D}=C^\infty_b(\mathbb{R})$. Condition $(E)$ is a part of the assumptions of theorem 3, condition $(N)$ is provided by item I) of  theorem 2. Let us check the Chernoff tangency: (CT1) follows from items IV) and III) of theorem 2, (CT2) is obvious from formula (2), (CT3) follows from item II) of theorem 2, (CT4) is a part of the assumptions of theorem 3. 

Therefore the statement of theorem 3 is true thanks to the statement of the Chernoff theorem and standard facts of the $C_0$-semigroup theory (see \cite{EN1} or shorter in section 3.4. of \cite{R1}).

\textbf{
\begin{center}
Discussion
\end{center}
}
\textbf{Remark 5.} Formula (2) can be rewritten in terms of generalized functions (=distributions) basing on the fact that $f(w)=\int_\mathbb{R}\delta(y-w)f(y)dy$: 
\begin{flushleft}
$(S(t)f)(x)=\int_\mathbb{R}\bigg[\frac{1}{4}\delta\left(y-x-2a(x)\sqrt{t}\right) + \frac{1}{4}\delta\left(y-x+2a(x)\sqrt{t}\right)+$ 
\end{flushleft}
\begin{flushright}
$+\frac{1}{2}\delta\big(y-x-2b(x)t\big) + tc(x)\delta(y-x)\bigg]f(y)dy.
$
\end{flushright}
Employing this equality one can rewrite (4) as a Feynman formula in which the integral kernel is a distribution (=generalized function).

\textbf{Remark 6.} One can use the presented method to solve equations of the type (1) not only with $\partial/\partial x$ and $\partial^2/\partial x^2$ but also with derivatives of any positive integer order: $\partial^3/\partial x^3$, $\partial^4/\partial x^4$ etc. To do this one needs to modify the formula for $S$ by taking more summands and tuning coefficients, allowing shifts proportional to $t, t^{1/2}, t^{1/3}, t^{1/4}, \dots$

\textbf{Remark 7.} It seems challenging to collaborate and compare numerically the presented method  with other novel  methods developed to solve parabolic equations \cite{Brad, BR, ZZCY}. The hypothesis is that a) our method will be faster on small times of evolution $t$ because it does not involve numerical integration or matrix inversion b) our method needs to take more approximation steps (greater $n$) for large $t$ as $t/n$ appears in the final formula (4).

\textbf{
\begin{center}
Acknowledgements
\end{center}
}
The author is thankful to D.\,V.~Turaev for the idea of employing shift operators for the role of $S(t)$ in the Chernoff theorem. It is also a pleasure to thank D.\,A.~Samsonov for the search of typos. Essential part of this work has been written in Lobachevsky Nizhny Novgorod State University and supported by the grant RNF 14-41-00044.

\textbf{
\begin{center}
References
\end{center}
}



\begin{thebibliography}{199}

\bibitem{EN1} K.-J. Engel, R. Nagel. One-Parameter Semigroups for Linear Evolution Equations. --- Springer, 2000.

\bibitem{Chernoff} Paul R. Chernoff, Note on product formulas for operator semigroups, J.  Funct. Anal.  vol. 2, no. 2, pp. 238-242, 1968. 

\bibitem{SWW} O.G. Smolyanov, H. von Weizsacker, O. Wittich. Chernoff's Theorem and the construction of semigroups, Proc. 7th Int. Conf. on Evolution Eq. and Their Main Areas of Appl. --- Evolution Equations: Applications to Physics, Industry, Life Sciences and Economics, vol. 55, pp 349-358,  Birkhäuser Basel, 2003.

\bibitem{STT} O.G. Smolyanov, A.G. Tokarev, A. Truman. Hamiltonian Feynman path integrals via the Chernoff formula. --- J. Math. Phys. 43, 10 (2002) 5161-5171.

\bibitem{SmHist} O.G. Smolyanov. Feynman formulae for evolutionary equations. --- Trends in Stochastic Analysis, London Math. Soc. Lect. Notes Series 353, 2009.

\bibitem{R1} I.D. Remizov. Solution to a parabolic differential equation in Hilbert space via Feynman formula - I. --- Model. and Anal. of Inform. Syst. 22, 3 (2015) 337-355. doi:10.18255/1818-1015-2015-3-337-355

\bibitem{Dubravina} V.A. Dubravina. Feynman formulas for solutions of evolution equations on ramified surfaces. --- Russ. J. Math. Phys. 2014, Vol. 21 (2) pp. 285-288.

\bibitem{SmSh} O.G. Smolyanov, N. N. Shamarov. Feynman Formulas and Path Integrals for Evolution Equations with the Vladimirov Operator. --- Proc. of the Steklov Math. Inst. 265 (1) 2009, 217-228.

\bibitem{R2}I.D. Remizov. Quasi-Feynman formulas --- a method of obtaining the evolution operator for the Schr\"odinger equation. --- J. Funct. Anal. vol. 270 iss. 12 (2016), 4540-4557 

\bibitem{Brad}A. Bradji. An analysis of a second-order time accurate scheme for a finite volume method for parabolic equations on general nonconforming multidimensional spatial meshes. --- Appl. Math. Comp., vol. 219, iss. 11, 6354-6371, 2013

\bibitem{BR}L. Bougoffa, R.C. Rach. Solving nonlocal initial-boundary value problems for linear and nonlinear parabolic and hyperbolic partial differential equations by the Adomian decomposition method. --- Appl. Math. Comp., vol. 225, 50-61, 2013

\bibitem{ZZCY} H. Zhang, Y. Zou, S. Chai, H. Yue. Weak Galerkin method with (r,r-1,r-1)-order finite elements for second order parabolic equations. --- Appl. Math. Comp., vol. 275, 24-40, 2016



\end{thebibliography}
\end{document}